\documentclass{svjour3}                     
\smartqed  
\usepackage{graphicx}
\begin{document}

\title{Asteroid Confusions with Extremely Large Telescopes}

\titlerunning{Asteroid confusions}    

\author{Gy. M. Szab\'o, A. E. Simon
}
\authorrunning{Gy. M. Szab\'o} 
\institute{   University of Szeged, Dept. Experimental Phiyics, 6720 Szeged D\'o{}m t\'er 9., Hungary \\
              Tel.: +36 544 052\\
              \email{szgy@titan.physx.u-szeged.hu}        
}

\date{Received: date / Accepted: date}

\maketitle

\begin{abstract}
Asteroids can be considered as sources of contamination of point sources and
also sources of confusion noise, depending whether their presence is detected in the image 
or their flux is under the detection limit. 
We estimate that at low ecliptic latitudes, $\approx$10,000--20,000 asteroids/sq. degree will be detected 
with an E-ELT like telescope, while by the end of Spitzer and Herschel missions,
infrared space observatories will provide $\approx$100,000 serendipitous asteroid detections.
The detection and identification
of asteroids is therefore an important step in survey astronomy. 
\end{abstract}

\section{Introduction}
\label{intro}

The presence of asteroids has been recognized to be a significant source of
confusion noise and contaminating point sources in images (e.g. Ivezi\'c et al. 2001, 
Tedesco \&{} Desert 2002; Meadows et al. 2004, Kiss et al. 2006, 2008). 
This confusion is most prominent in visible and infrared wavelengths, because 
asteroids reflect the sunlight in the visible, and have maximal thermal 
emission around 5--20 $\mu$m. As sky surveys go deeper, the number
of detected asteroids increases rapidly, and so do the number of asteroids near
the detection limit. 
In this paper we examine how the presence of asteroids deteriorate the quality
of data in sky survey images. We concentrate on three kinds of confusion:
by undetected asteroids, by unrecognized asteroids and by identified asteroids.
The following types of contaminations will be discussed in this paper:

\begin{figure*}
\includegraphics[width=\columnwidth]{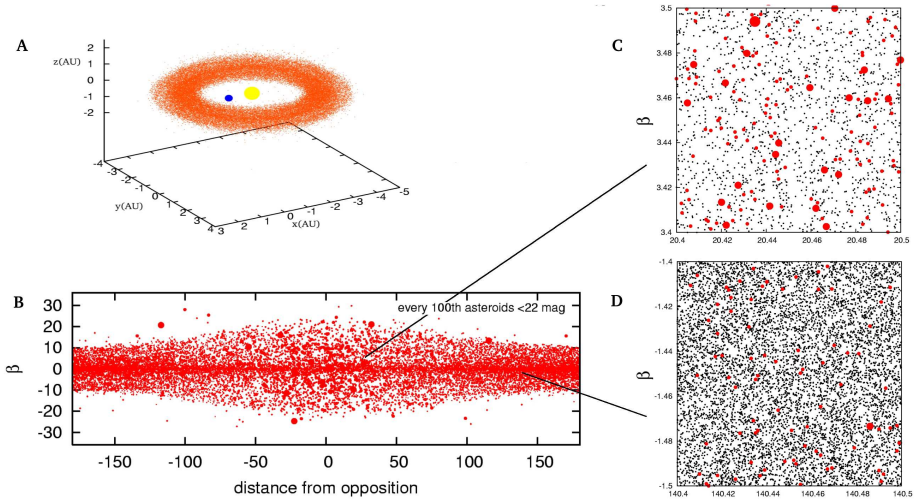}
\caption{The Solar System on 11 Sept. 2008. A: Space distribution of Statistical Asteroid
Model (SAM, every 30th asteroids are plotted), in yellow and blue are the Sun and Earth.
B: Distribution of asteroids $<$22 mag in the sky (every 100th asteroids are
plotted). C and D: Simulated ELT images with 1-min exposure taken at different ecliptical coordinates, 
(see the ecliptical coordinates of the areas on the axis labels) 
showing detected asteroids ($<$25 mag; red dots) and undetected/trailed 
asteroids (black dots).}
\label{fig:1}  
\end{figure*}

\begin{itemize}
\item{} By undetected asteroids.
The faint asteroid tracks contaminate the star field, and
covering a non-negligible fraction of the field 
they contaminate the precise photometry.
They must be considered as sources of confusion noise.
\item{} By detected sources not recognized as asteroids.
They essentially contaminate the star field, leading to potential
false candidates
for transient objects, variable stars, supernovae, GRB counterparts etc.
They especially contaminate the very automated measuring pipelines.
\item{} By asteroids recognized as asteroids.
They do not bother much unless they are blend with objects of interest.
\end{itemize}

\section{Estimating asteroid confusion in IR and visual wavelengths}
\label{sec:2}

Asteroids cause confusion because of the electromagnetic radiation
they reflect and emit. The IR thermal emission is related to the mean surface temperature, 
the geometric albedo, the thermal inertia and the 
visible cross section. Via the thermal inertia and thermal conductivity, 
the shape has also slight influence on the infrared flux. In the visual, we 
observe reflected light.
This is a product of the visible cross section and the geometric albedo,
scaled by a factor depending on the configuration geometry (distances and the
phase angle of the asteroid). In case of multicolor photometry, the knowledge of spectral albedo 
distribution is also necessary to predict the fluxes in each photometric bands.
Thus, for all confusion estimates, a necessary input is the 
size distribution function (SDF) of asteroids
in the Solar System. Determining the SDF is a non-trivial problem
while the high complexity of this challenge has been recognized recently (Parker et al. 2008).
An appoximate solution is to choose a global SDF for all asteroids. 
The most widespread model SDF for asteroids is the Statistical
Asteroid Model, SAM by Tedesco et al. (2005), which extends the empirical
global SDF of known asteroids into the $>$1 km range.

On the other hand, ELTs will observe much smaller
asteroids than the size limit of the SAM, and they will extend the detection limit to 
the 10 m -- 100 m size range. A 1-minute exposure with a 25~m-class telescope
enables us to do photometry of a 25 magnitude star with $S/N\approx 100$ quality.
The complete discussion of asteroid confusions needs extending the asteroid
brightness distribution model to $<30$ magnitude (i.e. a 30 mag asteroid
contaminates the photometry with 1 percent the flux of a 25 mag star, 
which is the predicted $S/N$).

Thus, the SAM needs extending toward small sizes,
in the 10 m -- 1 km range. For the following simulations, we designed an extension, and added 
many small asteroids to the SAM. The newly added point sources follows 
the same celestial distribution as the 1 km-sized asteroids, and they follow 
the same power slope as does the SAM towards its faint end (power index of $-2.75$).
This power is also consistent with the mean asteroid SDF from Sloan Digital Sky Survey, SDSS hereafter 
(Parker et al. 2008).

The results of our simulations illustrate how seriously asteroids can contaminate
the images toward specific ecliptical directions. In Fig. 1 we plotted the distribution
of SAM asteroids in the Solar System at the time of 11 Sep. 2008; Sun and Earth are also marked by yellow
and blue symbols. The celestial distribution of the SAM asteroids is plotted in panel B
(every 100th asteroids are plotted; the sizes of symbols are scaled to the brightness
of the individual objects). Panels C and D show the distribution of objects in our 
{\it extended} SAM. Red dots show asteroids that are brighter than 25 mag (which are likely
to be detected as point sources in e.g. an average E-ELT image), black dots
show objects in the 25--30 mag brightness range. Near the opposition point, at
3.5 degrees latitude, the number of detected asteroids will be $\approx 20,000$/sq.~degree,
and 2\%{} of the image area will be contaminated by asteroid trails. In comparison, 
at 40 degrees elongation from the Sun and at -1.5 degrees ecliptical latitude,
10,000 asteroids/sq. degree will be detected, however, the image area 
contaminated by asteroid trails will be as large as 10\%{}.
 
\subsection{Confusion noise by undetected asteroids}
\label{sec:2.1}

\begin{figure*}
\includegraphics[height=0.47\columnwidth]{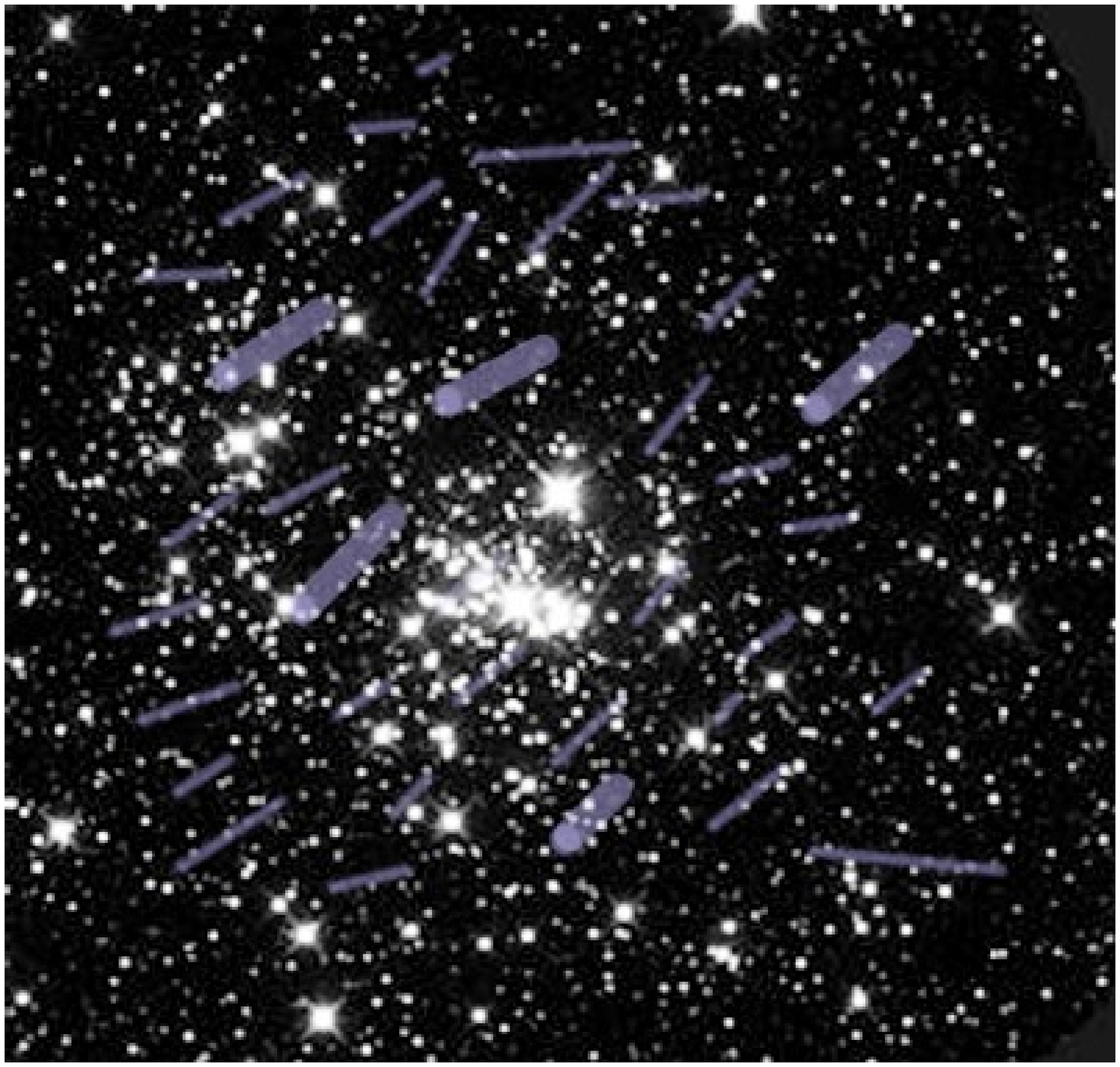}\hfill%
\includegraphics[height=0.47\columnwidth]{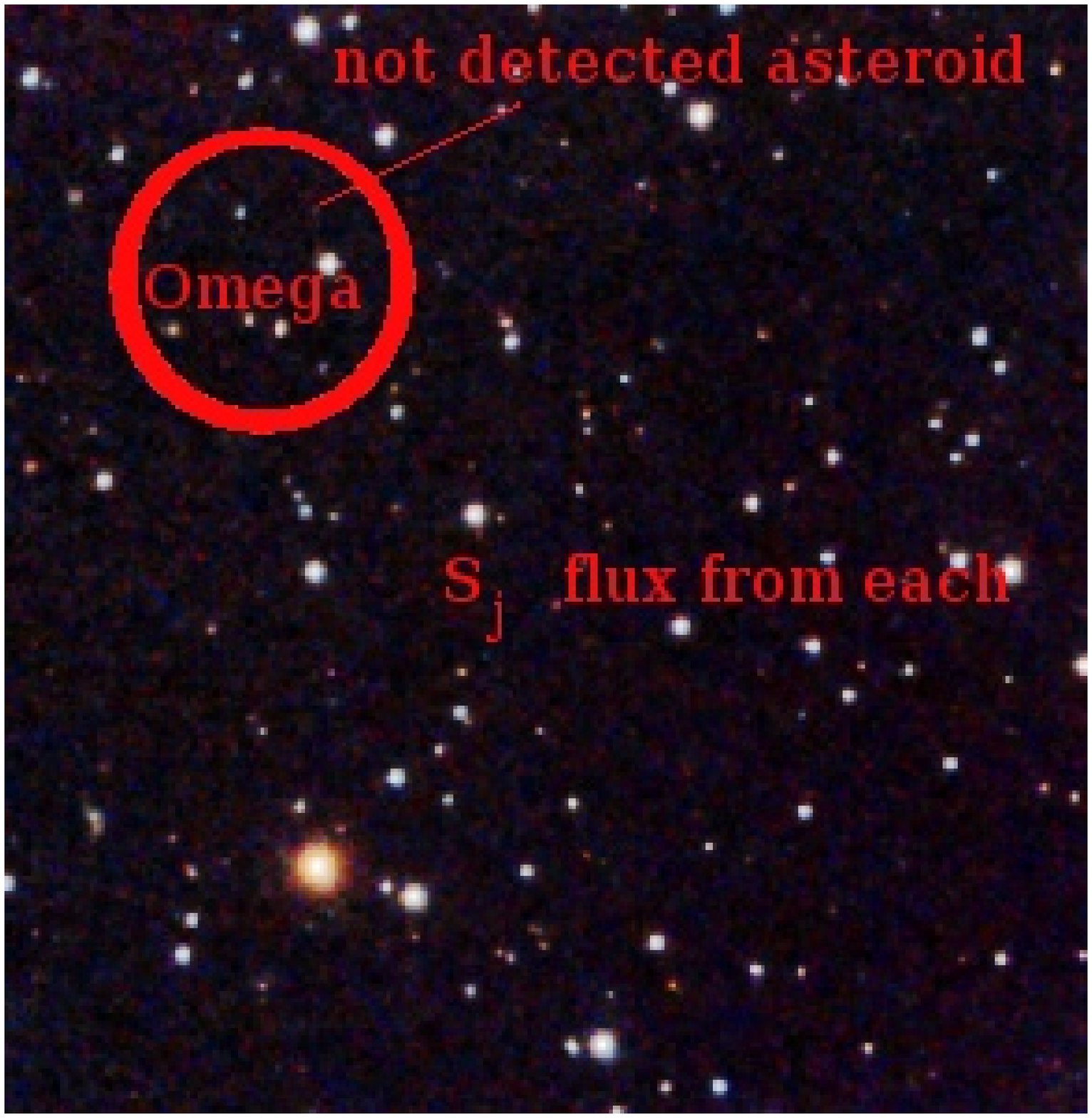}
\caption{Illustrations of confusion from undetected asteroids. Left panel: the moving faint
asteroids appear as faint ``noisy'' trails in the background, leading to confusion noise. The trails are
1--1.5 arcs/min long for typical main belt asteroids in opposition. Right panel: the
confusion noise is caused by the varying number of asteroid in counting cell
of a fixed size.}
\label{fig:1}  
\end{figure*}

According to Kiss et al. (2006, 2008), the asteroids must be considered as sources of 
confusion noise in IR. We summarize this following the cited papers.
Near the Ecliptic the effect of asteroids 
can be comparable to the contribution of Galactic cirrus emission 
and of the extragalactic background.
Assuming that the celestial distribution of asteroids is locally
Poisson-like, the $F_{lim}$ fluctuation power and the 
$\sigma_{lim}\left(\lambda_i,
S_{lim},  \Omega_p\right)$ confusion noise of undetected asteroids can be written as
$$\delta F_{lim}\left(\lambda_i, S_{lim} \right)= {1\over \Omega_c} \sum
_{S_i<s_{lim}} S_i^2 \left( \lambda_i \right),$$
$$\sigma_{lim}\left(\lambda_i, S_{lim},  \Omega_p\right) =
\bigg( \Omega_p \delta F_{lim}\left(\lambda_i, S_{lim} \right) \bigg) ^{1/2}.$$

Here $\Omega_c$ and $\omega_p$ are the effective solid angle of the counting
cell and the pixel of the instrument, respectively, and $S_j$ is the observed
flux of the asteroid at $\lambda_i$. The sum runs over all asteroids in the
counting cell (See Fig. 2 for illustration). With this technique, Kiss et al. (2006, 2008) presented an all-sky
map of the confusion noise by asteroids. Their conclusions were: 

\begin{itemize}
\item{} The confusion noise is most significant near the Ecliptic and 
peaks at the local anti-solar point. Seasonal variations were also detected.

\item{} Mid-infrared surveys like Spitzer/MIPS at 
24µm and Akari/IRC may be strongly affected by confusion noise in the vicinity 
of the ecliptic plane.

\item{} 3m-class IR telescopes like Herschel or 
SPICA will be unaffected. Asteroid confusion would not be negligible in 
anti-solar direction, however, solar aspect constraints for 
satellites usually do not allow to observe towards opposition targets.
\end{itemize}

A confusion noise estimator for several infrared instruments is hosted by the
Konkoly observatory \footnote{http://pc100.konkoly.hu/\~{}apal/sam/}. It was prepared to estimate the impact of the 
asteroids on infrared (IR) and submillimeter observations, 
from 5$\mu$m to 1000$\mu$m. The calculations are based on the 
Statistical Asteroid Model (Tedesco et al. 2005).

The Zodiacal light can be also considered as a possible error source in the infrared. 
\'Abrah\'am et al. (1997) mapped five ~0.5(deg) x0.5(deg) fields at low, intermediate and high ecliptic latitude 
at 25 $\mu$m with the photometer onboard the Infrared Space Observatory. 
According to their results, no structures were seen in these five sample fields. For an aperture of 3' 
diameter they found an upper limit for the underlying rms brightness fluctuations of +/-0.2\%{}, 
that corresponded at high ecliptic latitudes to +/-0.04 MJy/sr or +/-25mJy in the beam.
However, more sensitive instruments may detect the fluctuations of the zodiacal light,
and in the future it may show a non-negligible contribution to the far-IR confusion near the ecliptic plane
(Maris et al. 2006).

\subsection{Contamination by detected asteroids not recognized as asteroids}
\label{sec:2.2}

An asteroid is detected when it is identified as a point source in
at least one imaging bandpass.
If this happens, automated pipelines tend to add this detection to the list of
point sources, which can lead to severe consequences.
Primarily, these asteroids will contaminate the list of star fields in the visual wavelengths
because their photometric colors are near to the solar values. In the infrared, they may contaminate
the point sources with excess IR light (post-AGB candidates, debris disk candidates etc.)
due to their thermal radiation with temperatures $\approx$100--250 K.
When the objects move fast enough to produce deblended images in different
passbands,  it will result in false candidates for objects with nonstellar colors. 
In particular, the candidate quasars selected for SDSS spectroscopic 
observations would be significantly contaminated because they are recognized 
by their nonstellar colors (Ivezi\'c et al. 2001, Richards et al. 2001).

\begin{figure}
\includegraphics[width=0.67\columnwidth]{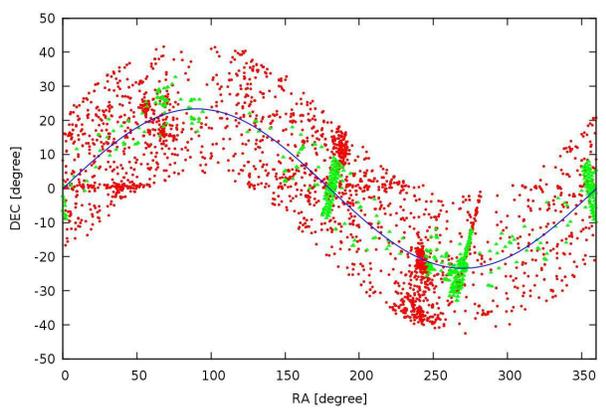}
\caption{Area covered by Spitzer MIPS scans (in red) and individual MIPS images
(in green) near Ecliptic. All images cover 830 sq. degree in total. We identified
8472 asteroid detections by serendipity in these images.
}
\label{fig:spitzermap}  
\end{figure}
\begin{figure*}
\includegraphics[width=0.47\columnwidth]{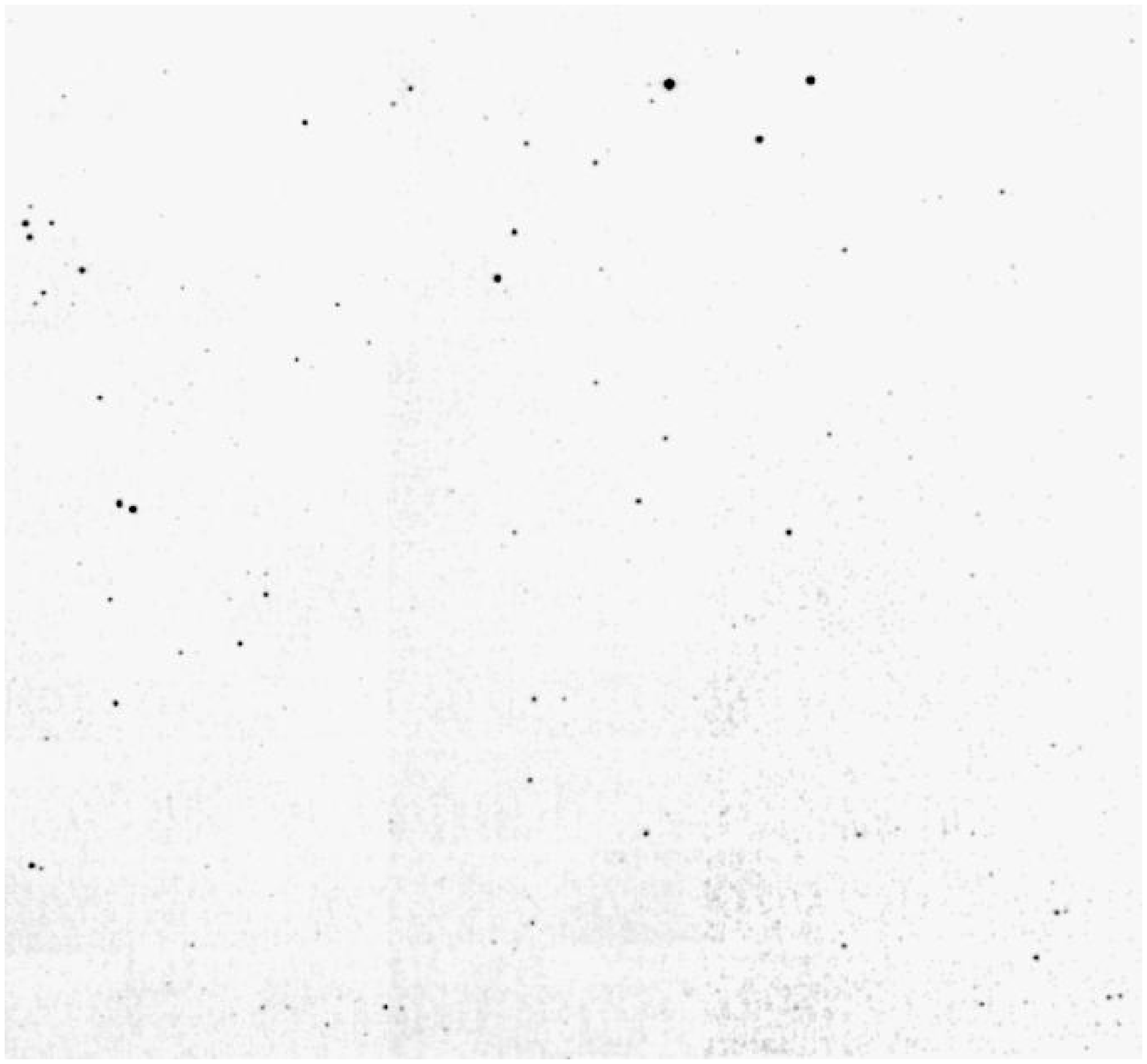}\hfill%
\includegraphics[width=0.47\columnwidth]{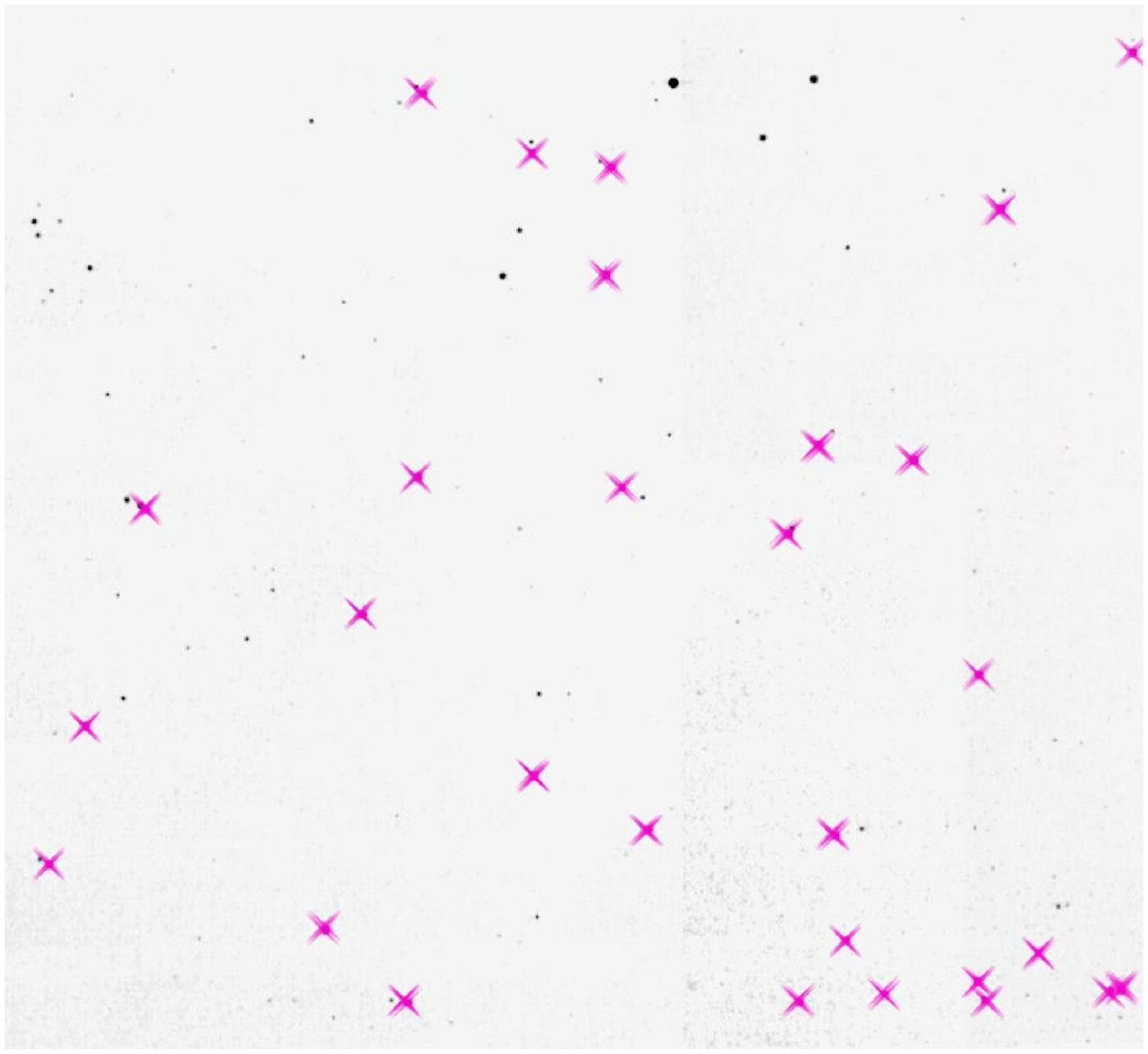}
\caption{Detections of asteroid in a sample Spitzer image (showing an area in Taurus,
Prop. num. 30816, P.I. Deborah Padgett) by serendipity.
Left panel: a region from the mosaic (300$\times$300 arc minute). Right
panel: the same image, but with predicted positions of known asteroids
(ephemeris generated by JPL/Horizons system). About 75\% of the predicted
positions can be linked to a point source detection in the Spitzer image
(see details in the text).
}
\label{fig:spitzer}  
\end{figure*}

\subsubsection{Case Study 1: Serendipitous asteroid detections by Spitzer}

To test how many main-belt asteroids can be identified in infrared space survey images serendipitously,
we tried to identify the known asteroids in selected Spitzer images. 
First, we collected image data for
MIPS scans near the Ecliptic. The images 
covered 830 sq. degrees in total (Fig. 3).  Then we configured
\verb xephemdbd \footnote{http://www.clearskyinstitute.com/xephem}~to identify the known asteroids in (and in the
close neighborhood of) these fields. While
\verb xephemdbd ~does not take all perturbations into 
account, we uploaded the list of candidate detections to the JPL Horizons Ephemeris 
Generator\footnote{http://ssd.jpl.nasa.gov/?horizons} and extracted the coordinates
for the time of observation. Finally we marked these positions onto the images (Fig. 4).

In total, we got a list of 8472 objects in 830 sq. degree sky area, 
that is $\approx$10 candidate asteroid detections$/$sq.
degree. Most of these asteroids can indeed be identified in Spitzer images as point sources.
Half of the detected asteroids can be found within a pixel to the predicted position. 
About 50\% of them is 3-4 arc seconds off, mostly because the 
orbits of these asteroids are not well known. Despite of this difference in position,
the identification looks to be certain for most of the objects. 20\%{} of the detected
objects is an entry in the SDSS Moving Object Catalog, and cross-linking the data
will result in an optical-infrared multi-wavelength catalog for more than 1500 asteroids.

Our prediction is that Spitzer will finally make $\approx$ 80--100 thousand detections
of known asteroids and the majority of them will have an optical counterpart in SDSS 
Moving Object Catalog (MOC).
These data will give key information about the structure of the asteroid belt, the formation
and evolution of our Solar System and extrasolar systems, too (e.g. Hines et al. 2007, 
Ryan et al 2008, Bhattacharya et al. 2008, Barucci et al. 2008, Lamy et al. 2008 etc.). 
The first results of SDSS MOC show the power of survey approach in Solar System studies 
(see e.g.: asteroid SDF and fine structure of families:
 Ivezi\'c et al. 2002, Parker et al. 2008; albedo variegation on asteroids: Szab\'o et al. 2004;
space weathering as family age indicator: Nesvorny et al. 2005; evolution of the shapes to less elongated
forms in 1-2 billion years: Szab\'o and Kiss, 2008; distribution of Trojan asteroids Szab\'o et al. 2008, Trojan
subfamilies: Roig et al. 2008).

A serious limit of Spitzer images is that post Basic Calibrated Data (post-BCD) 
images cannot be used for asteroid photometry, 
mainly because the image combination process more or less eliminates the moving objects from the
results (Z. Balog, personal communication). The other problem is the poor fit of predicted
and measured astrometric positions at least in a considerable fraction of cases (also noted
by Trilling et al. 2008). Therefore, the recognition and extraction of all 
serendipitously
observed asteroids require special algorithms, and also better known orbits. Asteroid detections by Pan-STARRS 
will lead to precise orbit elements for most of Spitzer asteroids, which will allow the precise
calculation of the ephemerides. It is likely that a full Spitzer asteroid catalog of 
$\approx$ 100 thousand detections can be released
in the post-Pan-STARRS era. Concerning the Herschel observatory, we can predict a similar
number of asteroid detections by serendipity in the MID-IR images (however, the numbers
highly depend on the fraction of MID-IR observations, which is likely to be low in the favor of FAR-IR
observations).

\subsubsection{Case Study 2: Asteroid Confusions at ELTs}

The problems with asteroid confusion occur differently with ELTs.
The field of view of typical E-ELT instruments will be of the order of 1 sq arcmin, maybe less. 
Furthermore, E-ELTs will likely work in the IR only since they intend to reach diffraction 
limit (which is of the order of 25 marcs for a 25m telescope). 
A surface density of 20000 asteroids/sdeg compares to about 5 asteroids per typical field of view. 
Certainly, trailing of the asteroid images will be of concern, however, this number
of asteroid detections will not be a serious source of error.

One can ask whether TNOs contaminate the images. The complete answer is difficult while we even do not know
the number and size distribution of TNOs. Kiss et al (2008) examined this question for a single TNO
object and compared its infrared flux to a similarly large main-belt asteroid. While TNOs are
$\approx$10 times farther from the Sun than MB asteroids, their black body equilibrium temperature is $\sqrt{10}$
times less. (The surface temperature is even lower when the asteroid has large optical albedo.) Consequently,
the IR radiation power is $10^2$ less for a TNO than for a similar MB object, which leads to a $10^4$ less 
flux density through the Lambert-law. Because of this factor, the TNOs do not seem to be severe sources of
confusion.

Interestingly, TNOs may somewhat deteriorate the fast photometric series of point sources by their occultations.
Because of diffraction effects the point sources do not disappear during the event, the occultation means a 
slight diminishment of the background object up to 50\%{} of the optical flux. This takes for $<\approx 50$ ms,
and was observed for several times by Georgevits (2006) and Roques et al (2008). The observed probability of 
one such event is in the order of $<\approx$1\%{} per night for a single star at low ecliptic latitudes.

\section{Asteroid data bases -- potentials with extremely large telescopes}
\label{sec:3}

We conclude that asteroid detection and identification
is a necessary step in reducing data from large telescopes, especially
those ones which work (also) in survey mode with highly automated pipelines.
The presence of the undetected and/or unidentified asteroids is a considerable error source
for many kind of observation (mostly those ones that utilize photometry of point sources, surface photometry and
stellar statistics). The solution is to detect as many asteroids 
as possible. In Solar System science,
this eliminated contamination will result in huge catalogs containing multicolor
photometry of asteroids. Moreover, catalogs from different sky surveys can be 
linked to each other, resulting in a multiwavelength spectral albedo
distribution of hundred thousands of asteroids. These data will finally provide basic information
about asteroids, e.g.  their size and albedo distributions, distribution of shape elongations,
and also detailed shape models, and finally will give a deep insight into the origin and evolution of the Solar
System.

Because asteroid confusion predictions are model-dependent, the limitations of a
statistical asteroid model (e.g. SAM) propagate to the predictions based on this model. 
In SAM, the SDF of asteroid is unique independently of their other properties such as
the orbital elements. Most recently, SDF has been recognized to vary
significantly in different families, and being different from size distributions 
for background populations. The families also have sub-structure: the cores tend to host
a larger fraction of large asteroids. In old families, a well-defined change of slope 
is shown that can be modeled as a broken power-law (Parker et al.
2008). Collisional evolution models predict very steep SDF 
for the small fragments in certain asteroid families (Michel et al. 2004).
The SAM does not take the small ($<$1~km) asteroids into account. One can
extrapolate an existing model (e.g. SAM) toward small sizes (just like we did in this paper), 
but may be that the SDF slope varies in that size range and the results will be misleading. 
These recently recognized SDF structures alert that all models which intend
to describe the SDF of all asteroids in the Solar System needs to be
updated.

The conclusion is that new statistical asteroid models are urgently needed for more precise 
predictions. The optimal model must be complete to at least the $>$100 m size range, 
while the SDF of different families and of the background should be adjusted separately.
However, this composition requires a large set of variables (thermal properties, 
spectral albedo  distributions and position-dependent SDFs for all the included families 
and the background) which by now have not been constrained by observations too well. The
infrared space observatories (e.g. Herschel) and the giant Earth-based telescopes
(e.g. LSST, E-ELT) will play a key role in building up a statistical asteroid model.
This will finally lead us to the better understanding of the asteroid confusions, and
not least, to the better understanding of the evolution of the Solar System.

\begin{acknowledgements}
This work has been supported by the Bolyai J\'anos
Research Fellowship of the Hungarian Academy of Sciences and the Hungarian OTKA Grant
K 76816. The traveling and living expenses were financed by the National Office for 
Research and Technology, Hungary (Mecenatura Grant) and a funding provided by the LOC. 
We thank Attila Mo\'o{}r for his kind help. ITM Dr. J. U.
\end{acknowledgements}

\end{document}